\documentstyle[aps,preprint]{revtex}
\begin{document}
\title{Influence of strings with axionic content on the
polarization of extragalactic radio sources}
\author{Luis Masperi}
\address{Centro At\'{o}mico Bariloche and Instituto Balseiro \\
Comisi\'on Nacional de Energ\'{\i}a At\'{o}mica\\
8400 Bariloche, Argentina}
\author{Sandra Savaglio}
\address{Dipartimento di Fisica, Universit\`a della Calabria\\
Arcavacata di Rende, 87036 Cosenza, Italy}
\maketitle
\begin{abstract}
It is shown that the axion concentrated in electroweak strings
which are consequently stabilized may produce a rotation of the
polarization angle of radiation from extragalactic sources larger
than that caused by a background field, and that axionic walls
attached to global strings have an effect which depends on the
frequency.  We discuss the validity of the classical and quantum
treatments of radiation and indicate that the latter produces
conversion of linear into circular polarization.  We describe
possible anomalies in the observations which might suggest these
influences of the axion.
\end{abstract}
\newpage
\section{Introduction}
One of the last pieces which would complete the standard model of
fundamental interactions is the axion, considered as the most
elegant explanation for the absence of strong CP violation$^{(1)}$.
 This particle would be a quasi-Goldstone boson caused by the
spontaneous breaking of a global U(1) symmetry at an energy scale
higher than the electroweak one.  Its mass, due to instanton
effects, is bounded in a window determined by astrophysical observations and
cosmological arguments$^{(2)}$, giving to  the axion the possibility of
being a candidate of CDM.
\par The interaction of the axion with the electromagnetic field
affects differently right and left circularly polarized radiation so
that the observation of extragalactic sources might give evidence
for its existence.  The consideration of background axion-like fields,
corresponding however to extremely light quasi-Goldstone bosons, has
allowed to establish bounds to their coupling with the
electromagnetic field$^{(3)}$.
\par In the present work we analyze the interaction
of radiation with static condensates of axions either in the form of
global strings, possibly stabilized by electroweak components, or
of electroweak strings with axions in their core.  For the latter
case it is reasonable to think that the axions in the string core
are very light both giving more stability to the string and making
the treatment of the interaction with electromagnetic field more
reliable.  The coherent nature of axionic condensates in strings
may produce a rotation of the polarization angle of radiation
larger than for a background field.
\par In Section II we describe the classical interaction of
radiation with static axionic field developing a perturbative
treatment,  which may be applied to rapid variations of the latter,
and propose electroweak strings with axionic content.  In Section
III we indicate that the adiabatic quantum approximation reproduces
the classical treatment and how the perturbative scattering of
photons on a static axionic condensate may produce the conversion
of linear to circular polarization.  In Section IV we suggest
possible astrophysical observations which might give evidence of
the above effects.
\section{Classical interaction of radiation with strings}
\par The interaction of a very light pseudoscalar field
$\sigma$ of the axion type with the electromagnetic field is
\begin{equation}
L_{\inf} = - g \sigma \vec{E} \cdot \vec{B}
\end{equation}
\par For time-independent $\sigma$, small coupling constant $g$ and
$\vec{\nabla} \sigma$ perpendicular to the electromagnetic fields
$\vec{E}$ and $\vec{B}$, Eq.~(1) gives way to modified Maxwell equations
\begin{eqnarray}
\Box \vec{E} = - g \nabla_i \sigma \nabla_i \vec{B} \nonumber \\
\Box  \vec{B} = g \nabla_i \sigma \nabla_i \vec{E}
\end{eqnarray}
which, defining $\vec F_{\pm} = \vec {E} \pm i \vec{B}$, are
equivalent to
\begin{equation}
\Box  \vec{F}_{\pm} = \pm i g \nabla_i \sigma \nabla_i \vec{F}_{\pm}
\end{equation}
\par If the variation of $\sigma$ is slower than that of the
electromagnetic field allowing $\vec{\nabla} \sigma$ to be taken as
a constant, the two circularly polarized waves $\vec{F}_{\pm}$
travel with different velocities$^{(4)}$, so  that a linearly
polarized wave traversing a width $L$ of the medium where the axion
field changes in $\Delta \sigma$ will emerge  with a polarization
angle rotated in
\begin{equation}
\Delta \theta = {1 \over 2} g \Delta \sigma
\end{equation}
\par Considering $\sigma$ as a background field, and being the
quasi-Goldstone boson of a symmetry broken at energy $f_{PQ}$, it
has been estimated$^{(3)}$ that the change of phase $\Delta
\sigma/f_{PQ} \approx \pi$.  In this way from the experimental
uncertainty of $\theta$ it is possible to determine a bound for the
coupling constant $g \simeq N \alpha/\pi f_{PQ}$, where $\alpha
\simeq 1/137$ and $N$ integer.
\par The application of Eq.~(4) seems more appropriate for static
coherent configurations of solitonic type.
\par The simplest one is the global string where the field $\psi(x)
= \exp (i \sigma(x)/f_{PQ})$ behaves as a vortex for large
distances from an axis for massless axion.  If two incident
linearly polarized electromagnetic rays pass perpendicularly to the
string axis on each side of its core, according to Eq.~(4) the
variation of their polarization angle will be of opposite sign.  If
the global string may act as a gravitational lens due to its large
mass the two images will have a difference in the polarization angle $\Delta
\theta \approx
g \pi f_{PQ}$ which is of a few degrees for small N.
\par On the other hand when the axion mass is taken into account
the configuration requires walls attached to the string separating
regions where $\sigma$ corresponds to different minima of the
potential$^{(5)}$.  The walls attract one another finally leaving
just one across which $\sigma/f_{PQ}$ changes in $2 \pi$.  This
variation of the axionic field is produced in the wall width
$\varepsilon \simeq 1/m_a \approx 1 cm$ for $m_a \simeq 10^{-5}
eV$.  Therefore for radio frequencies the approximation which led
to Eq.~(4) is no longer valid and a different treatment is required.
\par For that purpose we build a perturbative method starting from Eq.~(3).
The zeroth order approximation is
\begin{equation}
\Box \vec{F}_{\pm}^{(0)} = 0~~~,~~~ \vec{F}_{\pm}^{(0)} =
\vec{f}^{(0)}_{\pm} \exp [i(k_0 z - \omega t)],~~~ \omega = \mid k_0 \mid
\end{equation}
such that if $f_+^{(0)} = f_-^{(0)}$ we have linear polarization.
\par If we consider the known case in which the variation of $\sigma$ is
small in a wave-length, the first-order equation is
\begin{equation}
\Box \vec{F}_{\pm}^{(1)} = \mp g \ell k_0 \vec{F}_{\pm}^{(0)}~~~,~~~
\ell = {\partial \sigma \over \partial z}
\end{equation}
whose Fourier transformation gives
\begin{equation}
\vec {f}_{\pm}^{(1)} (k) = \mp g \ell k_0 \frac{\delta(k -
k_0)}{k^2 - \omega^2} \vec {f}_{\pm}^{(0)}
\end{equation}
The difference of the variation of velocities for the right and left polarized
waves defined by $\omega - k$ averaged with their Fourier component gives
\begin{equation}
{1 \over 2} \left [ \int dk (\omega - k) \frac{ f^{(1)}_+
(k)}{f^{(0)}_+} - \int dk (\omega - k)\frac{ f^{(1)}_-
(k)}{f^{(0)}_-}\right ] = {1 \over 2} g \ell
\end{equation}
which multiplied times the path $L$ reproduces Eq.~(4) checking the
validity of the perturbative procedure.
\par Passing to the opposite case where the change $\Delta
\sigma$ is produced through a thin wall, narrow in
comparison with the e.m. wave-length, the first-order equation is
\begin{equation}
\Box \vec{F}_{\pm}^{(1)} = \mp g k_0 \Delta \sigma \delta(z)
\vec{F}_{\pm}^{(0)}
\end{equation}
with the Fourier component
\begin{equation}
\vec{f}^{(1)}_{\pm} (k) = \mp { g \over 2 \pi} k_0 {\Delta \sigma
\over k^2 - \omega^2} \vec{f}_{\pm}^{(0)} ~~~ ,~~~ \omega = k_0
\end{equation}
The physical effect of this perturbation comes from the average of
$\omega - k$ as above of modes with positive $k$ for observations beyond the
wall
integrating up to the natural cut-off $\wedge = {2\pi \over
\varepsilon}$ which produces a rotation of the polarization angle
\begin{equation}
\Delta \theta = {g \over m_a \lambda_0} \Delta \sigma \ln (1+ m_a
\lambda_0) ~~~,~~~ \lambda_0 = 2 \pi/k_0
\end{equation}
This shows a dependence of the effect on the wave-length at
variance with the case of slowly varying $\sigma$.
\par A way to obtain larger effects on the polarization is to
include the axion field in an electroweak string.  This$^{(6)}$
corresponds to a vortex of the Higgs field $\varphi$ compensated by
the field $Z_{\mu}$ of the neutral vector boson.  Though this
configuration is metastable and even classically unstable$^{(7)}$
for realistic values of the Higgs mass, its stability can be
increased by including the field $\sigma$ in a potential with two
minima: the true vacuum corresponding to the phase with broken
symmetry where $\mid \varphi \mid = v~ \sigma = 0$ to avoid the
strong CP violation and a false one for the symmetric phase existing
inside the string where $\varphi = 0~ \sigma = \sigma_0$.  This last
non-vanishing value comes from the absence of one of the
contributions which must be compensated by the axion field, i.e.
$arg~ det M$ with $M$ mass matrix$^{(8)}$.  The local minimum
decreases the string energy making its formation easier.  On the
other hand the fact that $\sigma/f_{PQ}$ is a phase allows
excitations where its variation is  $n 2 \pi$ along a closed string
giving a quasi-topological reason for stabilization$^{(9)}$.  This
is somehow similar to the superconducting cosmic strings$^{(10)}$
but with the difference that there the excitations correspond to
charged Goldstone-bosons coming from the breaking of
$U(1)_{em}$ in the core whereas in our case $\sigma$ inside the
string may vary because the axions are lighter than outside, adjusting the
potential parameters, for being related to a higher-temperature phase.
\par The string energy per unit length
\begin{equation}
E = \int d^2 x \left [(\vec{D} \varphi)^* \cdot \vec{D} \varphi +
{1 \over 2} \vec{\nabla} \sigma \cdot \vec{\nabla} \sigma + V
(\varphi, \sigma) + {1 \over 2} (\vec{\nabla} \times \vec{Z})^2
\right ]
\end{equation}
where $\vec{D} = \vec{\nabla} - i g' \vec{Z}$ and a possible
potential$^{(9)}$ for $N = 1$
\begin{equation}
V = h ( \mid \varphi \mid^2 - v^2)^2 + \left [ m^2_a f^2_{PQ} + K (
\mid \varphi \mid^2 - v^2) \right ] (1 - cos { \sigma \over f_{PQ}})
\end{equation}
may be estimated in thin-wall approximation
\begin{equation}
E = a R^2 + b R + {c \over R^2}
\end{equation}
with $a$ determined by the difference of the two minima of
potential, $b$ by the changes in the interface and $c$ by the
Z-magnetic flux.
\par For the minimization of Eq.~(14) with respect to the core
radius $R$ one must
fix the numerical values of the potential parameters.  For a not
too high Higgs mass it is reasonable to take $h = 1$ together with
the electroweak breaking scale $v = 250 GeV$. To give the expected
axion mass in the broken symmetry phase $f_{pQ} \sim 10^{12} GeV$.  K is
related to
the axion mass inside the core
\begin{equation}
m'^2_a = K {v^2 \over f_{PQ}^2} - m^2_a
\end{equation}
One extreme position is i) to assume that when the core size is
maximum the two minima of the potential are almost degenerate which
occurs for $K = v^2/2$.  But in this way $m'_a \simeq 10 eV$ which
does not seem sensible and requires larger energy to allow the
oscillations of $\sigma$ along $z$.  The opposite choice
corresponds  ii) to imagine that $m'_a < m_a$ because the core has
a higher symmetrical phase.  In this case $K < 10^{-8} GeV$ and the
oscillations of the core radius along z are very gentle.  In any
case due, to the fact that $\sigma$ oscillates inside the core, for
an average situation we will take  $b$ as depending only on the gradient of
$\varphi$ in
the interface whose width will have as upper bound $1/m_a$.
\par With the option i) the core radius may be as large as $0.1pc$
with a linear energy density of $10^{31} g pc^{-1}$ for a
 $Z-$magnetic field $B \simeq 10^{-1} G$.  With the option ii) the
second term of Eq.~(14) becomes negligible compared with the first
one and different almost constant core sizes are possible, e.g. $R
\simeq 10^{-4} cm$, obviously with smaller $\varepsilon$, with $E \simeq
10^{20} g cm^{-1}$ and an
extremely high $Z$-magnetic field $B \simeq 10^{24} G$.
\par It is clear that the above numerical analysis is not
necessarily realistic since it is based on the toy-model Eq.~(13).
Anyhow the general idea is that an electromagnetic wave which
travels inside a large part of the string length, apart from the
change of the polarization angle in the interface given by
Eq.~(11), will suffer an additional rotation according to Eq.~(4)
which might be larger than that produced by a background field, due
to the coherent increase of $\sigma$ along the path.
\section{Quantum considerations}
When the interaction Eq.~(1) is taken as a source of quantum
changes, one must distinguish between the adiabatic approximation
according to which the photon adapts itself to the slow spatial
changes that it encounters in the axionic medium, and the sudden
transition to a different state.
\par In the first case one expects to obtain a result similar to
the classical one which depends on the gradient of $\sigma$. Considering as
interaction Hamiltonian
\begin{equation}
H_{\inf} \simeq g \int d \vec {r} \sigma \vec {E} \cdot \vec {B}~~~,
\end{equation}
since the modification of the electromagnetic conjugate momentum is
of order $g$, this interaction will change the quantum state of the
photon as it moves through  an axionic medium.  Taking the photon
as localized in a volume $\lambda^3$ over which the integration  of
Eq.~(16) must be performed, at the beginning $\sigma = 0$ to have
linear polarization e.g. along $\vec{e}_1$.  After the elementary path
$\lambda = t$ the photon encounters a constant $\sigma$ slightly
different from the initial one and, through the Schr\"{o}dinger
evolution operator, $H_{int}$ produces a component of polarization
$\vec{e}_2$ in the quantum state with amplitude
\begin{equation}
{\cal A} = g {\sigma \over 2} \vec {e}_1 \cdot(\vec{k} \times \vec{e}_2) t
\end{equation}
The rotation angle for the polarization of the photon may be defined as the
expectation value of an
operator $\vec{e}_1 \times \vec{e}_2$ which will receive
contribution only from the crossed term in the product of the final
states, giving as  result precisely Eq.~(4).  When the change of
$\sigma$ appears only in a width $\varepsilon < \lambda$ the
amplitude for the appearance of the polarization different from the
initial one receives contribution for $t \sim \varepsilon$ so
that, using Eq.~(17) as a rough estimation, a rotation inversely
proportional to $\lambda$ appears resembling Eq.~(11).
\par If instead of considering electromagnetic waves travelling
through an axionic medium we have the case of radiation polarized
along $\vec{e}(\vec{k})$ impinging on a localized microscopic distribution of
static $\sigma$, the perturbative treatment for sudden transitions
to different states applies.  Therefore, using Eq.~(1), the
cross-section for production of photons with polarization
$\vec{e}~'(\vec{k}')$ is
\begin{equation}
{d \sigma \over d \Omega} = {g^2 \over 16 \pi^2} (\vec{e}~' (\vec{k}') \cdot
(\vec{k} \times \vec{e} (\vec{k})))^2 (\sigma(\vec{q}))^2 k^2
\end{equation}
where $\sigma (\vec{q}) = \int d \vec {r} \sigma(\vec{r}) e^{i
\vec{q} \cdot \vec {r}}~~~,~~~ \vec{q} = \vec {k} - \vec{k}'.$
If $\sigma \sim \sigma_0$ constant in the interaction volume ${\cal V}$,
the maximum contribution of Eq.~(18) will be for $\vec{k} =
\vec{k}'$ producing a possible conversion of linear to circular photon at the
rate
\begin{equation}
{d \sigma \over d \Omega} \mid_{\vec{q} = 0} = {g^2 \over 16 \pi^2}
\sigma^2_0 {\cal V}^2 k^4
\end{equation}
which rises quickly with the frequency.
\par One may note that if $\sigma$ were a dynamical field the
cross-section for photon scattering would result of order $g^4$ but
this case is not realistic since the De Broglie wave-length of the
axion is$^{(2)}$ of order $10^4 cm$.
\par Regarding the strings with axionic content, the simple global
ones in terms of the field $\psi$ are finally attached to a domain
wall and decay quantum mechanically$^{(11)}$.  But if they include
electroweak components with two Higgs doublets their
stability increases due to conservation of $Z$-magnetic
flux$^{(12)}$. Since the dynamics of axionic strings
indicates$^{(13)}$ that they would radiate mainly just before the
QCD transition the earlier electroweak mechanism might stabilize a
number of them.  As for the electroweak strings which attract axions in their
core
according to Eq.~(12-14), apart from the increase of classical
stability given by the local potential minimum and the
quasi-topological argument, complete quantum stability may be
reached if axions are much lighter inside the core than outside
it$^{(14)}$. But this is possible for the original electroweak
strings which survive till almost the QCD transition when the axion
becomes massive.
\section{Possible relation with observations}
Normally the polarization angle of radiation from extragalactic
sources is at $0^0$ or $90^0$ with respect to the symmetry axis of
the source and is rotated by Faraday effect giving a quadratic
dependence on the wave-length for the observed angle$^{(15)(16)}$.
However the statistical analysis of radiogalaxis with redshift $z >
0.4$ shows$^{(15)}$ a possible excess in the distribution around
$-40^0$ which, in case of being real, might be attributed to
condensations of axionic field which affect some lines of sight.
\par Other observations which should be analyzed are those
corresponding to close images whose sources are considered known
so that their lines of sight are affected in a similar way by
{}Faraday effect.  This may happen for double-lobed sources which in
some cases show unexplained polarization differences$^{(17)}$.  It
would be analogously interesting to study the case of sufficiently
close sources.  As for the images of gravitational lenses, the
difference of polarization angle is normally explained by Faraday
effect$^{(18)}$, being possible to speculate about the existence of
examples where a global string acts as a lens.
\par In the emission of radiogalaxies there is a small percentage
of circular polarization$^{(17)}$ that in some abnormal cases
increases with the frequency and which might be attributed to a
conversion from linear to circular polarization by condensates of
axionic  matter.
\par There are also unexplained cases of polarization variations
with time for constant total intensity$^{(17)}$ which could be
eventually due to the oscillation of electroweak strings with
axionic content.
\par Finally in the optical region the dependence on the frequency
is not clear and the Faraday effect is excluded$^{(17)}$ so that
there might be room for axionic effects which, as we have seen, are
independent on the frequency or proportional to it in case of rapid
variation of axion field.\\
\\
\\
\\
\\
\\
\\
\\
\\
\\
\noindent{\bf Acknowledgments}
\par We are indebted to D. Harari and P. Veltri for interesting
discussions.  One of us (LM) thanks the Department of Physics of
the University of Calabria at Cosenza for the hospitality during
part of this work.  This research has been partially supported by
the grant PID~3965/92 of the Consejo Nacional de Investigaciones
Cient\'{\i}ficas y T\'{e}cnicas.
\newpage
\noindent{\bf References}
\begin{description}
\item{ 1)} R.Peccei and H.Quinn, Phys.Rev.Lett. \underline
{38}, 1440 (1977);Phys.Rev.D \underline {16}, 1791 (1977).
\item{2)} E. Kolb and M. Turner, The Early Universe (Addison
Wesley, Reading, MA, 1990).
\item{3)} D. Harari and P. Sikivie, Phys. Lett. B \underline
{289}, 67 (1992).
\item{4)} J. Harvey and S. Naculich, Phys. Lett. B \underline
{217}, 231 (1989).
\item{5)} P. Sikivie, Phys. Rev. Lett. \underline {48}, 1156 (1982).
\item{6)} T. Vachaspati, Nucl. Phys. B \underline {397}, 648 (1993).
\item{7)} M. James, L. Perivolaropoulos and T. Vachaspati, Phys.
Rev. D \underline {46}, R 5232 (1992).
\item{8)} G. Raffelt, Phys. Rep. \underline {198}, 1 (1990).
\item{9)} R. Fiore, D. Galeazzi, L. Masperi and A. M\'egevand,
Mod. Phys. Lett. A \underline {9}, 557 (1994).
\item{10)} E. Witten, Nucl. Phys. B \underline {249}, 587 (1985).
\item{11)} J. Preskill and A. Vilenkin, Phys. Rev. D \underline
{47}, 2324 (1993).
\item{12)} G. Dvali and G. Senjanovi\u c, Phys. Rev. Lett.
\underline {71}, 2376 (1993).
\item{13} R. Battye and E. Shellard, Axion string constraints,
Cambridge preprint (1994).
\item{14)} L. Masperi and A. M\'egevand, Stability of modified
electroweak strings, Bariloche preprint (1994).
\item{15)} S. Carroll, G. Field and R. Jackiw, Phys. Rev. D
\underline {41}, 1231 (1990).
\item{16)} A.Cimatti, S.Di Serego Alighieri, G.Field and R.Fosbury,
Astrophys.J.\underline {422},562 (1994).
\item{17)} D.Saikia and C.Salter, Annu.Rev.Astron.Astrophys.\underline {26},93
(1988).
\item{18)} A. Patnaik, J. Browne, L. King, T. Muxlow, D. Walsh
and P. Wilkinson, Mon.Not.R.Astron.Soc. \underline {261}, 435 (1993).
\end{description}
\end{document}